\documentclass[aps,prb,showpacs,tightenlines,twocolumn,nofootinbib,nobibnotes,superscriptaddress]{revtex4-1}
\usepackage{amsmath,amssymb,amsfonts,bm}
\usepackage{graphicx}
\usepackage{epstopdf}
\usepackage{dcolumn}
\usepackage{mathrsfs}
\usepackage{tgtermes}
\usepackage[colorlinks=true,linkcolor=blue,citecolor=blue, urlcolor=blue,bookmarks=false]{hyperref}
\bibpunct{[}{]}{,}{n}{}{}

\begin{document}
\title{Disorder-induced Majorana zero modes in a dimerized Kitaev superconductor chain}
\date{\today }
\author{Chun-Bo Hua}
\author{Rui Chen}
\author{Dong-Hui Xu}
\author{Bin Zhou}\thanks{binzhou@hubu.edu.cn}
\affiliation{Department of Physics, Hubei University, Wuhan 430062, China}

\begin{abstract}
Motivated by the recent experimental observation of the topological Anderson insulator in disordered atomic wires based on the Su-Schrieffer-Heeger (SSH) model, we study disorder effects on a dimerized Kitaev superconductor chain which is regarded as the superconductor version of the SSH model. By computing the real-space winding number and the zero-bias differential conductance, we analyze the topological phase transitions occurring in a dimerized Kitaev superconductor chain with disorder. It is found that disorder can induce a topologically nontrivial superconductor phase hosting Majorana zero modes (MZMs). We can regulate the appearance of disorder-induced MZMs by adjusting the dimerization parameter. Finally, we use the self-consistent Born approximation method to verify the numerical results.
\end{abstract}

\maketitle

\section{Introduction}
One of recent research interests in condensed matter physics is to realize topological superconductors (TSCs) with Majorana zero modes (MZMs) \cite{Read2000PRB,Kitaev2001,Wilczek2009NP,Service2011Science,Qi2011RMP,Alicea2012RPP,Leijnse2012SST,Stanescu2013JPCM,
Beenakker2013ARCMP,Elliott2015RMP,Sato2017RPP,He2018CSB}, which provide a platform for fault-tolerant quantum computation \cite{Kitaev2003,Stern2004PRB,Nayak2008RMP,
Akhmerov2010PRB,Hassler2010NJP,Sarma2015Majorana,Zhang2018TQC}. A well-known simplest toy model of TSCs is the Kitaev chain model \cite{Kitaev2001}, which describes a one-dimensional (1D) spinless $p$-wave superconductor chain that, under certain parameters, exhibits MZMs localized at its two ends. Up to present, various experimental suggestions have been projected to achieve 1D TSCs, including semiconductor-superconductor heterostructures \cite{Lutchyn2010PRL,Oreg2010PRL,Alicea2010Majorana,Leonid2012NP,
Anindya2012NP,MTDeng2012NanoLett,Mourik2012Signatures,
Finck2013PRL,Albrecht2016Nature,Gazibegovic2017Nature,Zhang2018Nature}, and magnetic atomic chain \cite{Stevan2014Science,Jeon2017Science,Michael2017NanoLett} or atomic ring with an external magnetic field \cite{LiJian2016NC} on the surface of $s$-wave superconductor.

In recent years, the Kitaev chain model has also been intensively investigated in the theoretical side. It is noted that Wakatsuki  \emph{et al}. \cite{Wakatsuki12014PRB} proposed a tight-binding model for hybrid system, known as the dimerized Kitaev chain model, consisting of the Su-Schrieffer-Heeger (SSH) model \cite{SSH1979PRB} and the Kitaev chain model \cite{Kitaev2001}. They studied the topological phase transitions of the dimerized Kitaev chain model by calculating the $k$-space winding number and the zero-bias differential conductance (ZBDC) \cite{Wakatsuki12014PRB}. Since then, a collection of studies on the dimerized Kitaev chain model has been reported \cite{Wang2017PRB,Ezawa2017PRB,Zhou2016CPB,Liu2016CPB,Zeng2016PRB,Wang2018PRE,Chitov2018PRB,Yu2019PRB,kobiaka2019arxiv}, such as the interacting dimerized Kitaev chain model \cite{Wang2017PRB,Ezawa2017PRB} and the quasi-1D dimerized Kitaev chain model \cite{Zhou2016CPB}.

On the other hand, the interplay between topology and disorder plays an important role in the recent research of topological matters and has been extensively investigated \cite{Brouwer2011PRL,Akhmerov2011PRL,DeGottardi2013PRL,Cai2013PRL,Rieder2013PRB,
Rieder2014PRB,Hui2014PRB,Adagideli2014PRB,Hui2015PRB,Zhou2015PRB,Cole2016PRB,Pekerten2017PRB,
Cai2017JPCM,Nava2017PRB,Awoga2017PRB,Liu2018PRB,
Monthus12018JPMT,Monthus22018JPMT,Habibi2018arxiv,Haim2019PRL}. Generally known, the topologically nontrivial phase is robust against weak disorder. When disorder is strong enough, the topologically nontrivial phase vanishes, and a topologically trivial phase appears. Interestingly, over the past decade, it has been found that disorder can switch a topologically trivial phase to a topologically nontrivial phase. The pioneering work of such disorder-induced topological phase is the discovery of the topological Anderson insulator \cite{Li2009PRL}. Since then, extensive theoretical studies of topological Anderson insulators have been carried out \cite{Jiang2009PRB,Groth2009PRL,Wu2016CPB,
Xing2011PRB,Guo2011PRB,ChenCz2015PRB,Orth2016SR,Su2016PRB,Kimme2016PRB,Chen2017PRBDirac,Chen2017PRBLieb,Guo2010PRL,Guo2010PRB,
ChenCz2015PRL,Shapourian2016PRB,Chen2018PRBWeyl,Chen2018PRBWeylFloquet,Titum2015PRL,Roy2016PRB}. Very recently, the experimental observation of the topological Anderson insulator has been reported in 1D disordered atomic chain based on the SSH model \cite{Meier2018Science} and optical lattices \cite{Stutzer2018Nature}. In addition, the disorder-induced topological phases in TSCs have also attracted much attention \cite{Borchmann2016PRB,Qin2016SR,Lieu2018PRB,Habibi2018PRB}. Analogous to the topological Anderson insulator, Borchmann \emph{et al.} proposed the concept of the Anderson topological superconductor, a disorder-induced topological state in superconductor systems \cite{Borchmann2016PRB}. Recently, Lieu \emph{et al.} studied disorder effects on Kitaev chain model with longer-range hopping and pairing terms, and presented the transformation of phase boundaries under the influence of disorder \cite{Lieu2018PRB}. Moreover, the combined effects of disorder and interaction in the Kitaev chain model have also been investigated by several research groups \cite{Lobos2012PRB,Gergs2016PRB,McGinley2017PRB,Thakurathi2018PRB,Kells2018PRB}.

Motivated by the recent experimental observation of the topological Anderson insulator in disordered atomic wires based on the SSH model \cite{Meier2018Science}, an intriguing question is whether the topological Anderson phase can occur in the dimerized Kitaev superconductor chain model which is regarded as the superconductor version of the SSH model. In this paper, we study disorder effects on a dimerized Kitaev superconductor chain. The dimerized Kitaev chain model with disorder belongs to the class BDI in the classification table \cite{AZ1997PRB,Schnyder2008PRB,Schnyder2009AIP,Ryu2010NJP,Chiu2016RMP}. In the clean limit, the dimerized Kitaev superconductor chain supports the topologically nontrivial and trivial phases depending on the model parameters. The topologically nontrivial phase is characterized by one pair of MZMs located at the ends of the chain. We focus on the topological properties of the dimerized Kitaev superconductor chain when Anderson-type disorder is turned on. We investigate the topological phase transitions by applying three different methods, which include the real-space winding number (RSWN), the ZBDC, and the self-consistent Born approximation methods. We uncover rich phase diagrams under the influence of disorder and find that the MZMs are stable for weak disorder but strong disorder takes MZMs away. Interestingly, based on the computations of the RSWN and the ZBDC, it is observed that a topologically nontrivial superconductor phase can be induced by disorder at a certain parameter values in the dimerized Kitaev superconductor chain, companied with the disorder-induced MZMs located at the ends of the chain. Finally, the self-consistent Born approximation method is used to confirm our numerical results for weak disorder.

The rest of the paper is organized as follows. In Sec.~\ref{Model}, we introduce a dimerized Kitaev chain model with disorder. Then, we give the details of numerical methods in Sec.~\ref{Methods}, and provide numerical results for studying the topological phase transitions of the system in Sec.~\ref{Numerical}. Subsequently, in Sec.~\ref{BA}, we confirm the numerical results in weak disorder by the self-consistent Born approximation method. Finally, we summarize our conclusions and discuss the experimental schemes of the system in Sec.~\ref{Conclusion}.

\section{Model}
\label{Model}
\begin{figure}[hptb]
	\includegraphics[width=8cm]{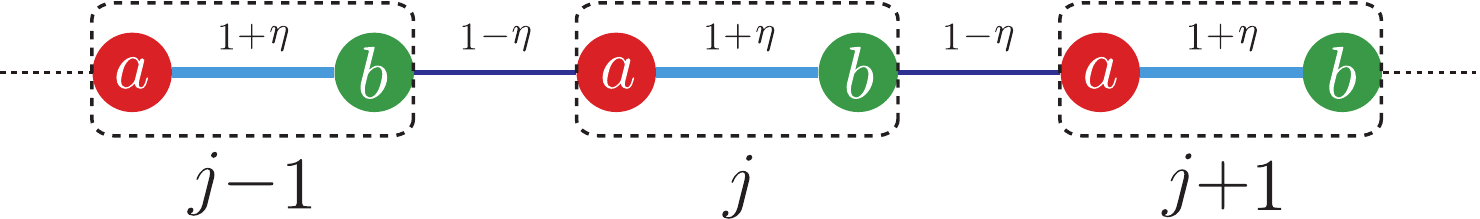} \caption{(Color online) Schematic illustration of the dimerized Kitaev superconductor chain. $j$ denote the $j$-th unit cell enclosed by the black dashed box, and a unit cell contains two sublattices marked by red (green) filled circle $a$ ($b$). The (blue thin and cyan thick) lines represent the dimerization of the particles, and  the thin (thick) lines indicate the intercell (intracell) couplings. The spatial differences ($1\pm\eta$) of dimerization parameter are shown.}%
\label{fig1}
\end{figure}

We start with the Hamiltonian of the dimerized Kitaev chain model \cite{Wakatsuki12014PRB} with Anderson-type disorder, and the illustration of the model is shown in Fig.~\ref{fig1}. Here we assume that the lattice cell number is $L$ and the lattice constant is equal to one. The Hamiltonian is written as
\begin{align}
H=&-\sum_{j=1}^{L}\mu _{j}\left(c_{a,j}^{\dag}c_{a,j}\!+\!c_{b,j}^{\dag}c_{b,j}\right)\nonumber  \\
&-t\sum_{j=1}^{L-1}\left[\left(1\!+\!\eta\right)c_{b,j}^{\dag }c_{a,j}
\!+\!\left(1\!-\!\eta\right)c_{a,j+1}^{\dag}c_{b,j}\!+\!\text{H.c.}\right]\nonumber\\
&+\!\Delta\sum_{j=1}^{L-1}\left[\left(1\!+\!\eta\right)c_{b,j}^{\dag }c_{a,j}^{\dag
}\!+\!\left(1\!-\!\eta\right)c_{a,j+1}^{\dag}c_{b,j}^{\dag}\!+\!\text{H.c.}\right],
\label{H}
\end{align}
where $j$ is the lattice coordinate, $a$ and $b$ denote the sublattice indices, $c_{a/b,j}^{\dag }$ ($c_{a/b,j}$) is the creation (annihilation) fermionic operator on site ($a/b$, $j$), $t$ is the hopping amplitude, and $\Delta$ is the strength of $p$-wave superconducting pairing. The dimerization parameter $\eta$ ($|\eta|<1$) is the space-dependent variable of hopping and pairing terms, and the spatial differences ($1\pm\eta$) of dimerization parameter are shown in Fig.~\ref{fig1}. The disorder term is $\mu _{j}=\mu +W\omega _{j}$, where $\mu$ is the chemical potential, $\omega_{j}$ is the uniform random variable chosen from $\left[ -0.5,0.5\right]$, and $W$ is the disorder strength. In subsequent calculations, the energy unit is set as $t$, and the chemical potential $\mu$ is fixed as $0$.

We discuss the topological class of the model based on the time-reversal ($T$), particle-hole ($R$) and chiral ($C$) symmetries. These three symmetry operators in the real space are defined as
\begin{equation}
T=K,R=(\tau _{x}\otimes I_{2L})K,C=\tau _{x}\otimes I_{2L},
\end{equation}
where $K$ is the complex conjugate operator, $\tau_{x}$ is the Pauli matrix acting on the particle-hole degree of freedom and $I_{2L}$ is a $2L\times2L$ unit matrix. The Hamiltonian (\ref{H}) satisfies the relations
\begin{equation}
THT^{-1}=H,RHR^{-1}=-H,CHC^{-1}=-H.
\end{equation}
Therefore, the Hamiltonian (\ref{H}) has the time-reversal, the particle-hole and the chiral symmetries, and belongs to the class BDI of the Altland-Zirnbauer classification table \cite{AZ1997PRB,Schnyder2008PRB,Schnyder2009AIP,Ryu2010NJP,Chiu2016RMP}. The class BDI is characterized by the $\mathbb{Z}$ index, and its topological invariant is the winding number in 1D TSCs.

\begin{figure}[hptb]
	\includegraphics[width=5.5cm]{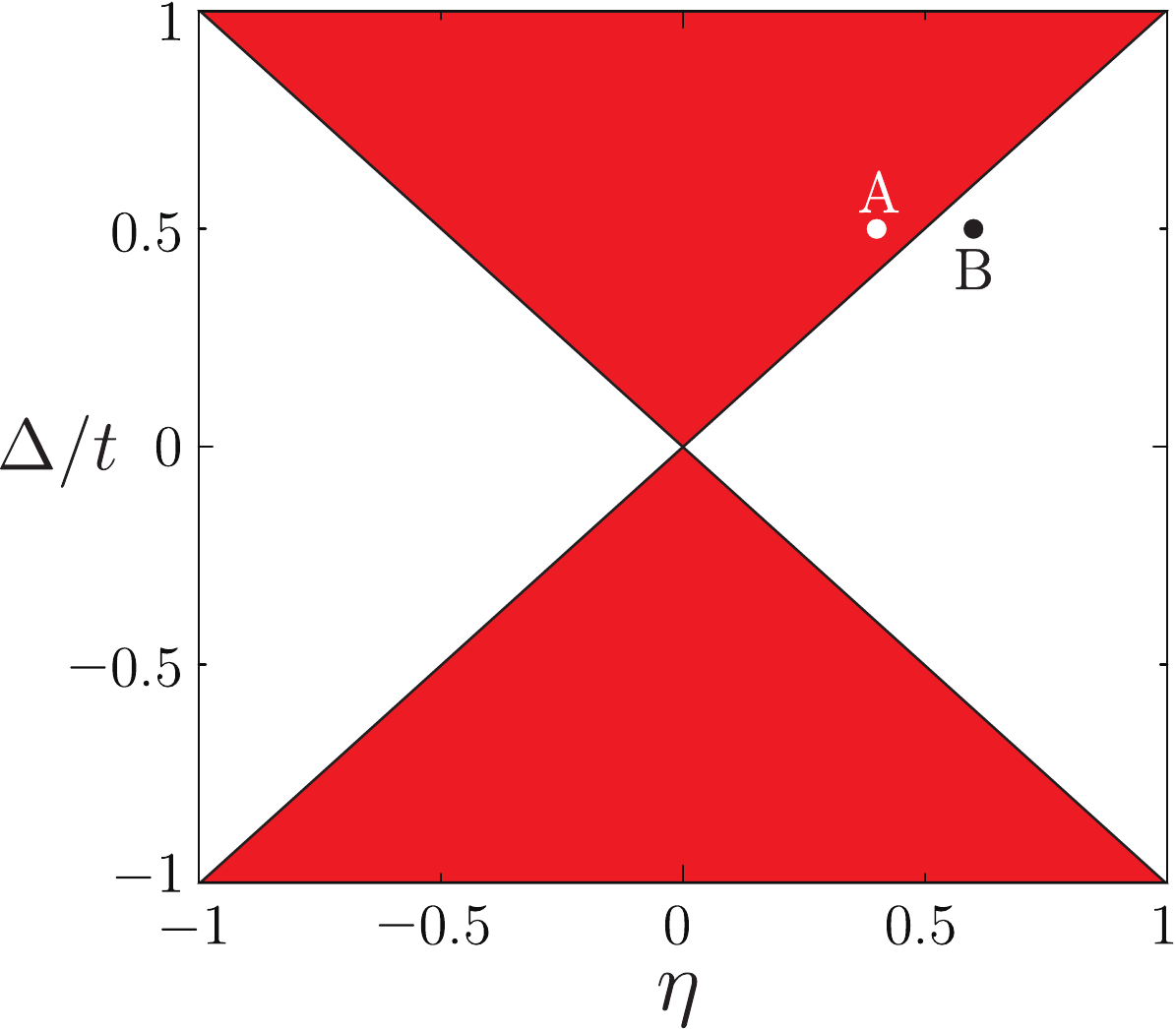} \caption{(Color online) Phase diagram of clean dimerized Kitaev chain model on the $\eta$-$\Delta/t$ plane. The red and white regions represent the topologically nontrivial and trivial phases, respectively. The white point A and black point B correspond to $(\eta,\Delta/t)=(0.4,0.5)$ and $(0.6,0.5)$, respectively. Here $\mu=0$.}%
\label{fig2}
\end{figure}

Before studying disorder effects on the system, we first review the phase diagram of the clean dimerized Kitaev chain model ($W=0$) [shown in Fig.~\ref{fig2}], which is obtained by calculating the $k$-space winding number \cite{Wakatsuki12014PRB}. The phase boundaries are $|\Delta/t|=|\eta|$. When $|\Delta/t|>|\eta|$, the phase is topologically nontrivial, and it indicates that there exists one pair of MZMs at the ends of the chain. But when $|\Delta/t|<|\eta|$, the phase is the topologically trivial superconducting phase, and there is no MZMs at the ends of the chain.

\section{Numerical Methods}
\label{Methods}
In this section, we introduce the RSWN and ZBDC methods to study disorder effects on topological phase transitions of the model. Similarly, the two methods can also be applied to systems without disorder.

The topology of 1D TSCs in class BDI is characterized by the winding number $\nu$. However, because the translational symmetry is broken in the disordered Kitaev chain, the $k$-space winding number in the frame of the Bloch wavefunctions is not applicable. Hence, the winding number must be handled in the real space. The topological invariant of the disordered Kitaev chain can be obtained by the RSWN method \cite{Prodan2016JFA,Mondragon2014PRL,Habibi2018PRB,Meier2018Science}. First, one constructs the homotopically equivalent flat band Hamiltonian: $Q=P_{+}-P_{-}$, where $P_{\pm}$ are the projection operators onto the positive or negative energy of the Hamiltonian $H$. And the projection operators $P_{\pm}$ are given by
\begin{equation}
P_{+}=\sum_{\epsilon _{j}>0}\left\vert \psi _{j}\right\rangle \left\langle\psi _{j}\right\vert, P_{-}=I_{4L}-P_{+},
\end{equation}
where $\left\vert \psi _{j}\right\rangle$ is the wave function of the $j$-th state with eigenvalue $\epsilon _{j}$ and $I_{4L}$ is a $4L\times4L$ unit matrix.

Next, one calculates the projection operators of the chiral symmetry operator $C$. The eigenvalues of $C$ are $\pm1$, and $C$ can be written as $C=C_{+}-C_{-}$, where $C_{\pm}$ are the projection operators of these eigenvalues. Then, the chiral-symmetric operator $Q$ can be decomposed as $Q=Q_{+-}+Q_{-+}$, where $Q_{+-}=C_{+}QC_{-}$ and $Q_{-+}=C_{-}QC_{+}$  \cite{Mondragon2014PRL}. The RSWN is given by
\begin{equation}
\nu =-\text{Tr}\left\{Q_{-+}\left[ X,Q_{+-}\right] \right\},
\label{nu}
\end{equation}
where $X$ is the position operator and "Tr" indicates a trace per number of unit cells $L$. The case with $\nu=0$ corresponds to the topologically trivial phase, and $\nu=1$ corresponds to the topologically nontrivial phase.

To check the result of the RSWN method, we also study transport properties of the disordered system. The setup is assumed that one semi-infinite normal metal lead is attached to one end of the superconductor chain. The normal metal lead Hamiltonian $H_{l}$ is described by Eq.~(\ref{H}) by installing $W$, $\Delta$ and $\eta$ to zero. The Hamiltonian $H_{ld}$ that represents the connection of normal metal and superconductor is described by Eq.~(\ref{H}) by installing $\mu$, $W$, $\Delta$ and $\eta$ to zero. Here, the three hopping amplitudes of superconductor, normal metal lead and the connection of normal metal and superconductor are all set to be equal.

In order to calculate the differential conductance of the normal-superconductor (NS) junction, we first calculate the scattering matrix $S$ of the NS junction by adopting the recursive Green's function method \cite{Ii2012PRB,Lewenkopf2013JCE,Liu2013PRB,Zhou2017PLA}. The scattering matrix $S$, related to the Green's function, is given by \cite{Lee1981PRL,Fisher1981PRB}
\begin{equation}
S^{\alpha\beta}=-\delta_{\alpha,\beta}+i\left[\Gamma^{\alpha}\right]^{1/2}G_{r}\left[\Gamma^{\beta }\right]^{1/2},
\end{equation}
where $\alpha$ and $\beta$ denote the electron ($e$) or hole ($h$)
channels. $S^{\alpha \beta }$ is an element of the scattering matrix and expresses the scattering amplitude of a outgoing $\beta $ particle caused by the incoming $\alpha $ particle. $\Gamma^{\alpha }=i\left(\Sigma^{\alpha}_{r}-\Sigma^{\alpha}_{a}\right)$ is the linewidth function of $\alpha $ particle, where $\Sigma^{\alpha}_{r/a}$ is the retarded (advanced) self-energy of $\alpha $ particle for the lead. The self-energy is $\Sigma_{r/a} =H_{ld}^{\dag}g_{r/a}H_{ld}$. And $g_{r/a}=\left[E_{l}\pm i0^{+}-H_{l}\right] ^{-1}$ is the retarded (advanced) Green's function of the semi-infinite lead, where $E_{l}$ is the fermi level of the lead and is set as $0$. $G_{r}$ is the retarded Green's function of the superconductor, and can be expressed as
\begin{equation}
G_{r}=\left[E+i0^{+}-H-\Sigma _{r}\right] ^{-1}.
\end{equation}
The physical meaning of the scattering matrix is: $S^{ee}$ denotes the normal electron reflection, and $S^{eh}$ denotes the local Andreev reflection.

The differential conductance of the NS junction as a function of electron incident energy $E=eV$ is represented by the scattering matrix \cite{Anantram1996PRB,Nilsson2008PRL,Liu2013PRB,Blonder1982PRB,Pientka2012PRL}
\begin{equation}
\frac{dI}{dV}=\frac{e^2}{h}\sum_{\alpha }\text{Tr}\left[1-%
\text{sgn}\left( \alpha \right) S^{\dag e\alpha }\left( eV\right)
S^{e\alpha }\left( eV\right) \right],
\label{dIdV}
\end{equation}
where $\text{sgn}\left(e\right) =1$, $\text{sgn}\left(h\right) =-1$. For the NS junction, the bias $V$ is applied at the normal metal lead and the superconductor is grounded. The ZBDC of the NS junction is calculated by $dI/dV$ at zero bias $V=0$. At the interface of the NS junction, the local Andreen reflection occurs \cite{Andreev1964JETP}. When the MZMs exist at the ends of the chain, the MZMs-induced resonant Andreev reflection occurs \cite{Law2009PRL}. The ZBDC is $2e^2/h$ if there is one pair of MZMs located at the ends of the chain, and $dI/dV=0$ if no MZMs \cite{Law2009PRL,Flensberg2010PRB}.

\section{Numerical Results}
\label{Numerical}
In this section, we numerically investigate disorder effects on the topological phase transitions of the dimerized Kitaev chain model. The topological phase diagrams with different parameters will be presented. First of all, based on the computation of the RSWN ($\nu$) and the ZBDC ($dI/dV$) as a function of the disorder strength ($W/t$), respectively shown in Eq. (\ref{nu}) and Eq. (\ref{dIdV}), we study disorder effects on the topological phase transitions at several parametric spatial points near the phase boundary of the clean phase diagram [marked by the white point A and black point B in Fig.~\ref{fig2}]. The corresponding model parameters of points A and B in Fig.~\ref{fig2} are $(\eta,\Delta/t)=(0.4,0.5)$ and $(\eta,\Delta/t)=(0.6,0.5)$, respectively.
\begin{figure}[hptb]
	\includegraphics[width=8.5cm]{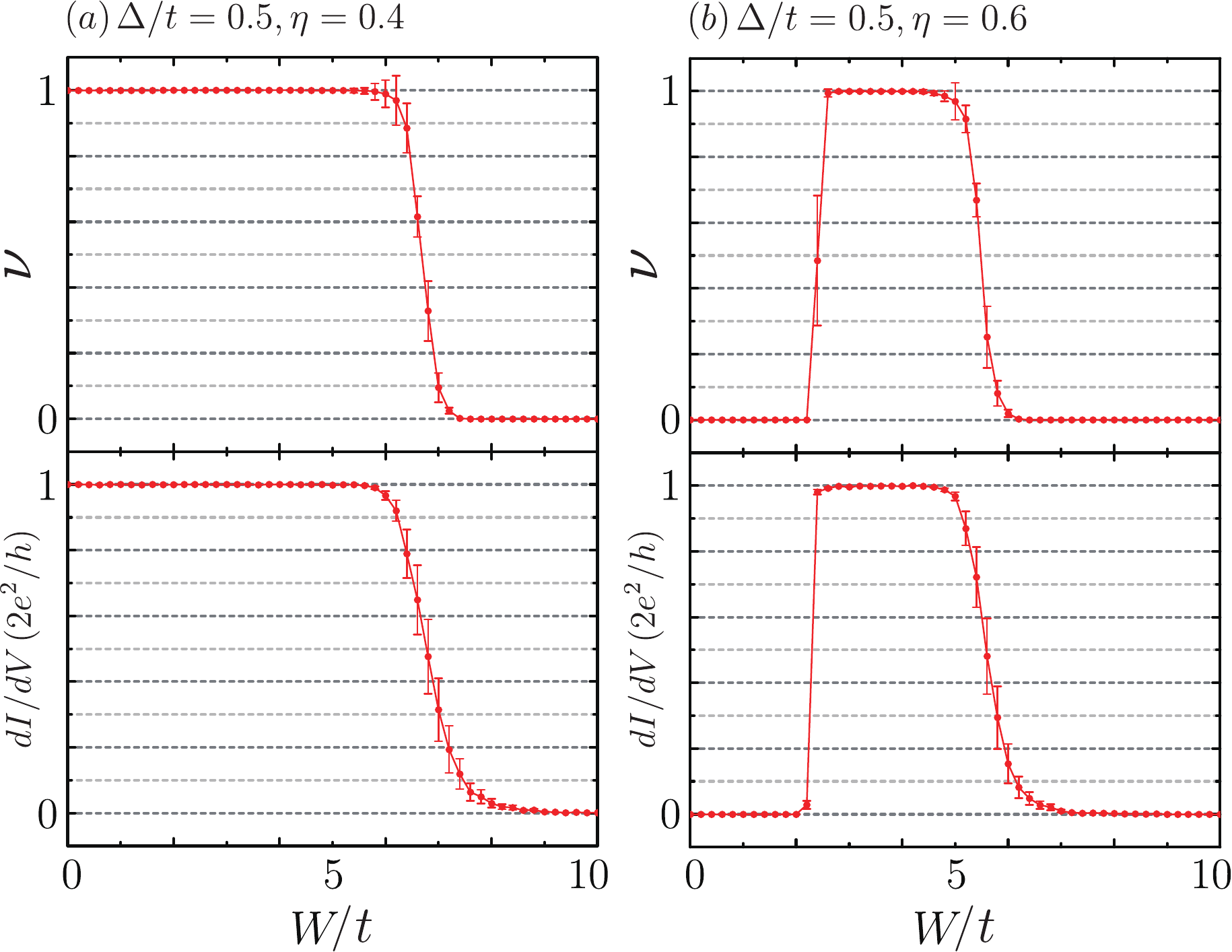} \caption{(Color online) The RSWN ($\nu$) and the ZBDC ($dI/dV$) as a function of the disorder strength for (a) $\eta=0.4$ and (b) $\eta=0.6$. We take the parameters $\Delta/t=0.5$ and $\mu=0$. In calculating the RSWN (the ZBDC), the size of the superconductor chain is taken as $L=500$ ($1000$), and the error bar indicates standard deviation of $500$ ($1000$) samples.}%
\label{fig3}
\end{figure}

The RSWN ($\nu$) and the ZBDC ($dI/dV$) of the two points as a function of the disorder strength ($W/t$) are shown in Fig.~\ref{fig3}. When $\eta=0.4$ and $\Delta/t=0.5$, the chain without disorder is the topologically nontrivial phase. With the disorder strength increasing, as shown in Fig.~\ref{fig3}(a), it is found that the topologically nontrivial phase remains stable in the case of weak disorder, which is characterized by $\nu=1$. Meanwhile, the ZBDC is $2e^2/h$, and the quantized conductance indicates the appearance of the MZMs-induced resonant Andreev reflection \cite{Law2009PRL}. Further increasing $W/t$, a topological phase transition occurs at $W/t=5.8$, beyond which both the RSWN ($\nu$) and the ZBDC ($dI/dV$) decay to zero, and the system is transformed into a topologically trivial phase.

For the point B ($\eta=0.6$ and $\Delta/t=0.5$) in Fig.~\ref{fig2}, the corresponding phase is topologically trivial in clean limit. When Anderson-type disorder is turned on, with the disorder strength increasing, as shown in Fig.~\ref{fig3}(b), it is interesting to observe that the RSWN changes from $\nu=0$ to $\nu=1$ at $W/t=2.6$ and return to $\nu=0$ at $W/t=4.8$. A plateau of the RSWN ($\nu=1$) maintains in a certain range of disorder strength ($2.6\leq W/t\leq4.8$). The plateau indicates that a topologically nontrivial phase is induced by disorder. A similar result can also be obtained by studying transport properties, and we find that the variation of the ZBDC with the disorder strength is similar to that of the RSWN. With the disorder strength increasing, the ZBDC jumps from $dI/dV=0$ to $2e^2/h$ at $W/t=2.6$, and goes back to $0$ at $W/t=4.8$. It is obvious that the conductance plateau can match well with the plateau of the RSWN. Thus, it means that in the dimerized Kitaev chain (with model parameters $\eta=0.6$, $\Delta/t=0.5$, and $\mu=0$) the MZMs can be induced by disorder when the disorder strength is in the region of $2.6\leq W/t\leq4.8$.

\begin{figure}[hptb]
	\includegraphics[width=8cm]{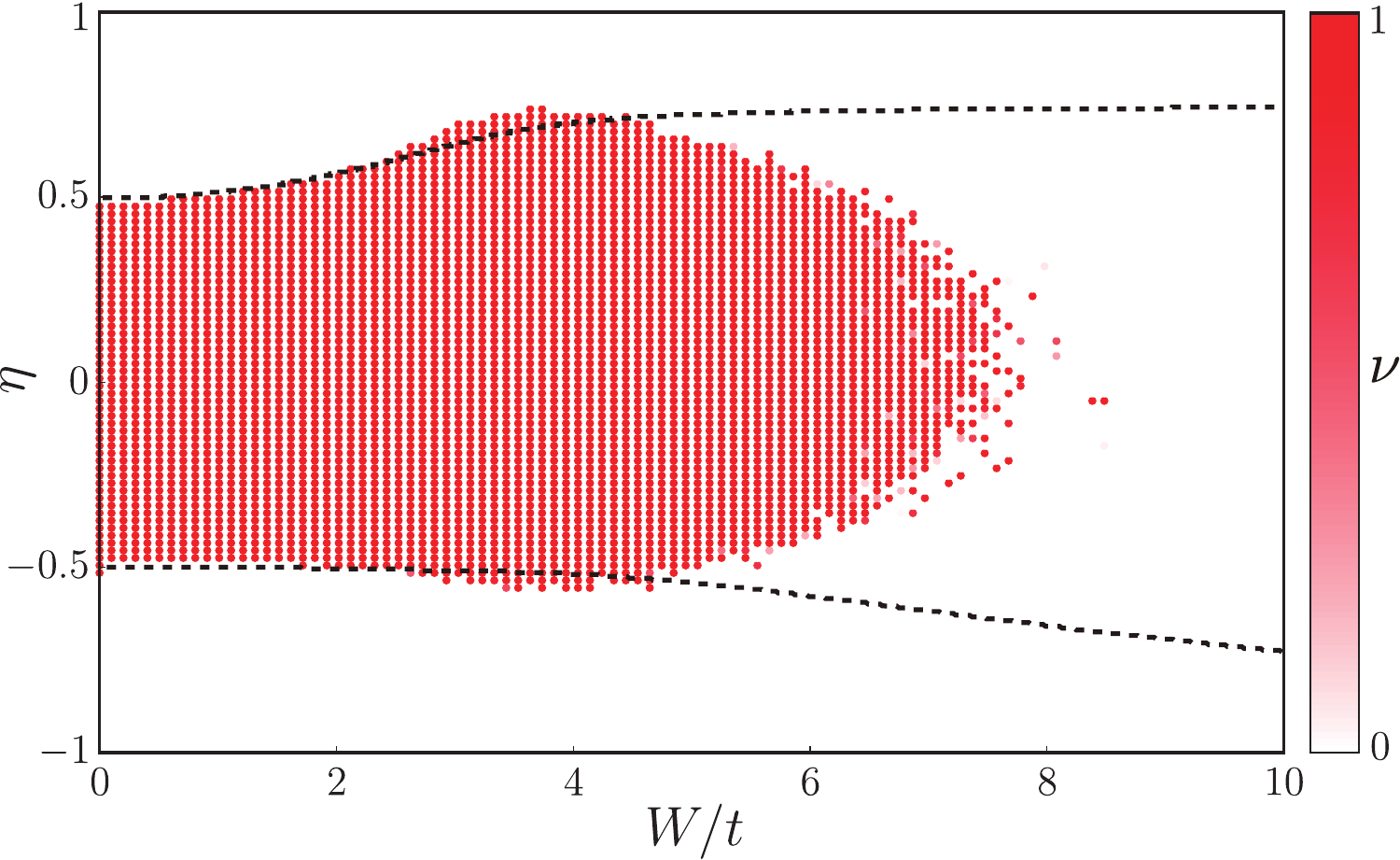} \caption{(Color online) Phase diagram in ($W/t$, $\eta$) space for the dimerized Kitaev chain with disorder obtained by calculating the RSWN. We take the parameters $\Delta/t=0.5$ and $\mu=0$. The red region denotes the topologically nontrivial phase ($\nu=1$), and the white region denotes the topologically trivial phase ($\nu=0$). The black dashed lines are determined by the self-consistent Born approximation method. The size of the superconductor chain is taken as $L=1000$. }%
\label{fig4}
\end{figure}
Additionally, the topological phase diagram for the dimerized Kitaev chain with disorder in the ($W/t$, $\eta$) space is plotted in Fig.~\ref{fig4}, where $\Delta/t=0.5$ and $\mu=0$. The color map shows the values of the RSWN $\nu$. It is necessary to point out that in numerically calculating the RSWN the size of the superconductor chain should be taken enough long to avoid the finite size effect of MZMs \cite{Zhou2011PRB,Chen2019PRB}. In Appendix~\ref{FSE}, we numerically investigate the finite-size effect in the dimerized Kitaev chain model in the clean limit. Here we take $L=1000$. Each point in Fig.~\ref{fig4} corresponds to a single realization of the disorder potential, which turns out to be sufficient for determining the region of the topologically nontrivial phase. The red region corresponds to the topologically nontrivial phase characterized by $\nu=1$, and the white region corresponds to the topologically trivial phase with $\nu=0$. In absent of the dimerization effect in the chain ($\eta=0$), figure \ref{fig4} shows that the topologically nontrivial phase remains stable up to the maximum disorder strength about $W/t\approx8$. While with the increasing of the dimerization parameter, the maximum disorder strength, beyond which the topologically nontrivial phase is destroyed, gradually decreases. And the disorder-induced topologically nontrivial phase regions in a range of parameters $W/t$ and $\eta$ are distinctly presented in the phase diagram shown by Fig.~\ref{fig4}.

\begin{figure}[hptb]
	\includegraphics[width=8cm]{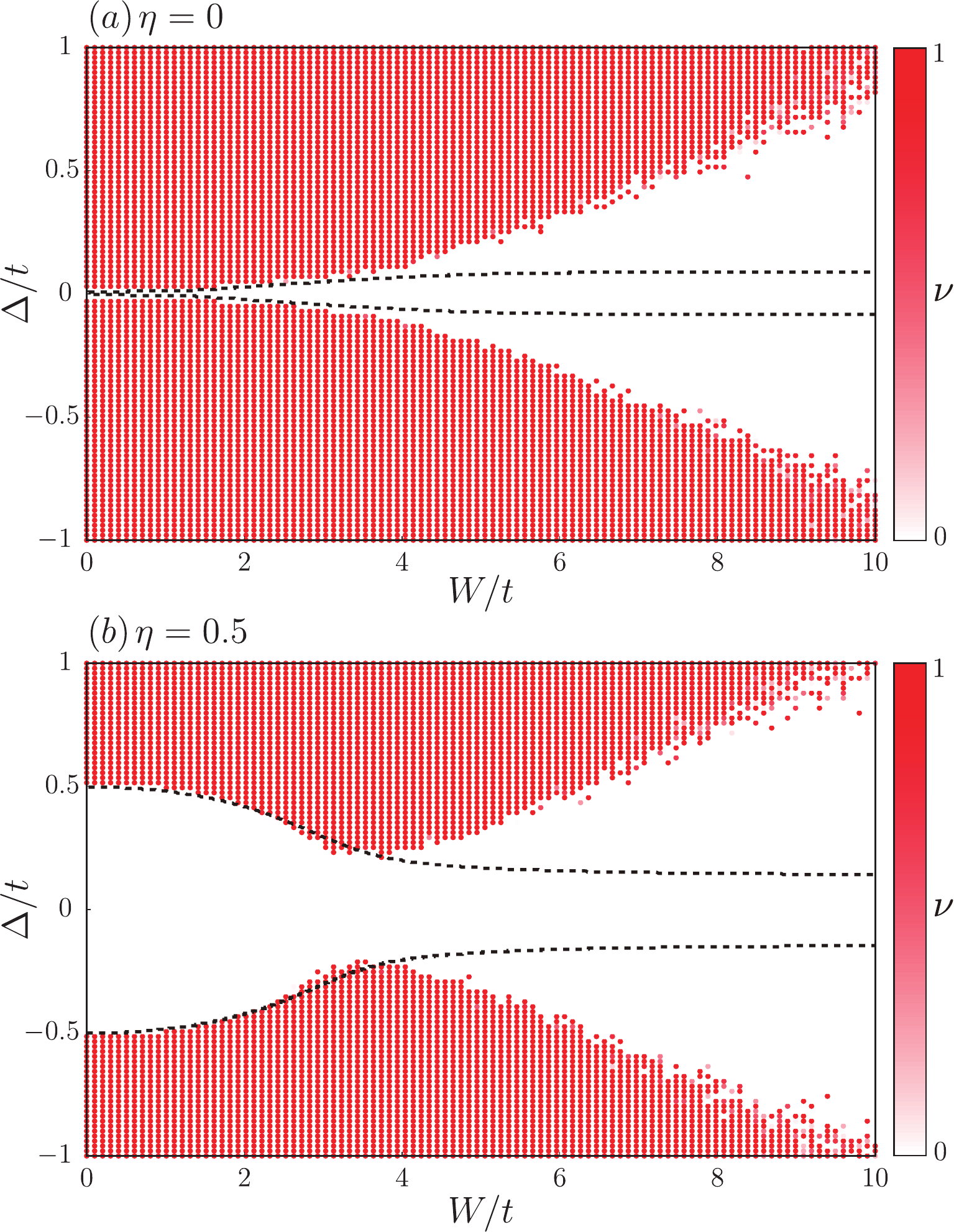} \caption{(Color online) Phase diagram in ($W/t$, $\Delta/t$) space for the dimerized Kitaev chain with disorder obtained by calculating the RSWN. We take the parameter (a) $\eta=0$ and (b) $\eta=0.5$. The red region denotes the topologically nontrivial phase ($\nu=1$), and the white region denotes the topologically trivial phase ($\nu=0$). The black dashed lines are determined by the self-consistent Born approximation method. The size of the superconductor chain is taken as $L=1000$. Here $\mu=0$.}%
\label{fig5}
\end{figure}
Figures ~\ref{fig5}(a) and ~\ref{fig5}(b) show two phase diagrams in the ($W/t$, $\Delta/t$) space with $\eta=0$ and $\eta=0.5$, respectively. When $\eta=0$, the dimerized Kitaev model returns to the original Kitaev model, and the phase diagram of the original Kitaev model in Fig.~\ref{fig5}(a) obtained by the RSWN method coincides with the phase diagram in the literature \cite{DeGottardi2013PRL} obtained by the transfer matrix method. In Fig.~\ref{fig5}(a), we find that the RSWN $\nu$ changes from $\nu=1$ to $\nu=0$ with increasing of the disorder strength. Further, in the case without the dimerization effect ($\eta=0$), it is shown that the larger the strength of $p$-wave superconducting pairing $\Delta$ is, the more robust against disorder the topologically nontrivial phase is. As shown in Fig.~\ref{fig5}(b), the dimerization effect suppresses the topologically nontrivial phase, however, dimerization and disorder have a combined influence on topological properties of the Kitaev chain. In the present of the dimerization effect ($\eta=0.5$), the disorder-induced topologically nontrivial phases are observed in Fig.~\ref{fig5}(b). In this case, the dimerized Kitaev chain with Anderson-type disorder becomes the the Anderson topological superconductor \cite{Borchmann2016PRB}.

\section{Self-consistent Born Approximation}
\label{BA}
Another approach that is often used to understand the phase transitions induced by disorder is the self-consistent Born approximation method. Through this method, the role of disorder can be regarded as a self-energy, and the disorder-induced self-energy can renormalize the model parameters.

In the clean limit, one introduces a periodic boundary condition, i.e., $c_{L+1}^{\dag}=c_{1}^{\dag}$, and uses the Fourier transform $c_{a/b,j}=\frac{1}{\sqrt{L}}\sum_{k}e^{ikj}c_{a/b,k}$ of the operator $c_{a/b,j}$, where $k$ is the wave vector and $-\pi<k\leq \pi $. Then, the Hamiltonian (\ref{H}) can be expressed in the Bogoliubov-de Gennes (BdG) form by introducing the four-component operator $C_{k}^{\dag}=\left(\begin{array}{cccc}c_{a,k}^{\dag },c_{b,k}^{\dag },c_{a,-k},c_{b,-k}\end{array}\right)$. The $k$-space Hamiltonian is \cite{Wakatsuki12014PRB}
\begin{equation}
H=\frac{1}{2}\sum_{k}C_{k}^{\dag }H_{0}\left( k\right) C_{k},
\end{equation}
with
\begin{equation}
H_{0}\left( k\right) =\left(
\begin{array}{cccc}
-\mu  & z & 0 & w \\
z^{\ast } & -\mu  & -w^{\ast } & 0 \\
0& -w & \mu  & -z \\
w^{\ast } & 0 & -z^{\ast } & \mu
\end{array}%
\right),
\label{H0}
\end{equation}
where $z\left( k\right)=-p_{+}-p_{-}e^{-ik}$, and $w\left( k\right)=-q_{+}+q_{-}e^{-ik}$, with $p_\pm=t\left(1\pm\eta\right)$ and $q_\pm=\Delta\left(1\pm\eta\right)$. In the $k$-space Hamiltonian (\ref{H0}), the dimerization parameter $\eta$ is coupled with the pairing strength $\Delta$ and the hopping amplitude $t$. Since the self-energy is independent of the momentum \cite{Groth2009PRL,Borchmann2016PRB,Qin2016SR}, the renormalization parameters are $\mu_{R}=\mu-\Sigma_{0}$, $p_{+R}=p_{+}-\Sigma_{1}$ and $q_{+R}=q_{+}+\Sigma_{2}$. Then, the self-energy can be expressed as
\begin{equation}
\Sigma=\Sigma_{0}\left(\tau_{z}\otimes\sigma_{0}\right)+\Sigma
_{1}\left(\tau_{z}\otimes\sigma_{x}\right)+\Sigma_{2}\left(\tau_{y}\otimes\sigma_{y}\right),
\end{equation}
where $\tau_{i}$ and $\sigma_{i}$ are the Pauli matrices acting on the particle-hole and the sublattice degrees of freedom, respectively. The disorder-induced self-energy in the self-consistent Born approximation method reads \cite{Groth2009PRL,Borchmann2016PRB,Qin2016SR}
\begin{align}
\Sigma=&\frac{W^{2}}{12}\frac{1}{2\pi }\int_{\text{FBZ}}dk\left( \tau _{z}\otimes \sigma _{0}\right)\nonumber\\
&\times\left[ \omega +i0^{+}-H_{0}\left( k\right) -\Sigma\right] ^{-1}
\left( \tau _{z}\otimes \sigma _{0}\right).
\label{Sigma}
\end{align}
This integration is over the first Brillouin zone (FBZ). Here $\omega$ is the frequency, and we employ $\omega=0$ with focusing on the static limit. Equation (\ref{Sigma}) is to be solved self-consistently. The derivation of the disorder-induced self-energy formula in the self-consistent Born approximation method is given in Appendix~\ref{Derivation}.

We numerically calculate the $k$-space winding number of the renormalized Hamiltonian $H\left( k\right) =H_{0}\left( k\right) +\Sigma$. First, we introduce a unitary transformation
\begin{equation}
U=\frac{1}{\sqrt{2}}\left(
\begin{array}{cccc}
1 & 0 & 1 & 0 \\
0 & 1 & 0 & 1 \\
-i & 0 & i & 0 \\
0 & -i & 0 & i%
\end{array}\right).
\end{equation}
Then the renormalized Hamiltonian is converted into a block-off-diagonal form with
\begin{equation}
UH\left( k\right) U^{\dag }=\frac{i}{2}\left(
\begin{array}{cc}
0 & A_{k} \\
A_{k}^{\dag } & 0
\end{array}%
\right),
\end{equation}
and
\begin{equation}
A_{k}=\left(
\begin{array}{cc}
-\mu_{R}  & z_{R}-w_{R} \\
w_{R}^{\ast }+z_{R}^{\ast } & -\mu_{R}
\end{array}%
\right),
\end{equation}
where $z_{R}=z+\Sigma_{1}$, and $w_{R}=w-\Sigma_{2}$.
The $k$-space winding number is given as
\begin{eqnarray}
\nu_k &\equiv &-\text{Tr}\int_{-\pi }^{\pi }\frac{dk}{2\pi i}A_{k}^{-1}\partial_{k}A_{k} \nonumber \\
&=&-\int_{-\pi }^{\pi }\frac{dk}{2\pi i}\partial_{k}\ln\left[\det\left(A_{k}\right)\right].
\end{eqnarray}
By numerically calculating the $k$-space winding number as functions of ($W/t$, $\eta$) and ($W/t$, $\Delta/t$), we obtain the black dashed lines in Fig.~\ref{fig4} and Fig.~\ref{fig5}, respectively. The black dashed lines denotes the phase boundary line of the $k$-space winding number between $\nu_k=1$ and $\nu_k=0$. In Fig.~\ref{fig4}, the region inside two black dashed lines is determined by $\nu_k=1$, while in Fig.~\ref{fig5}, the region inside two black dashed lines corresponds to the case with $\nu_k=0$. It is found that the results based on the self-consistent Born approximation method can match well with the numerical ones in the case of weak disorder. However, it is also observed that there exists disagreement for strong disorder. The reason for the mismatch in strong disorder is that the weak scattering potential approximation is used in the derivation of the disorder-induced self-energy formula in the self-consistent Born approximation method (see Appendix~\ref{Derivation}). Therefore, the self-consistent Born approximation method can only be applied to the case of weak disorder.

\section{Conclusion and discussion}
\label{Conclusion}
In this paper, we investigate the topological phase transitions of a dimerized Kitaev chain model with Anderson-type disorder. To determine the topological phase of the system, we numerically calculate the RSWN and the ZBDC of the finite chain, and observe a phase transition from a topologically trivial phase to a topologically nontrivial phase hosting MZMs located on the ends of the chain at a finite disorder strength. We present the phase diagrams based on the numerical results of the RSWN as functions of the disorder strength and the dimerization strength (the superconducting pairing strength), and it is shown that the interplay between dimerization and disorder has an interesting influence on topological properties of the Kitaev chain. Finally, we find that the result obtained by the effective medium theory based on the self-consistent Born approximation can confirm the numerical results of the RSWN for weak disorder.

Recently, there are several experimental schemes to achieve 1D TSCs, such as the semiconductor-superconductor heterostructures \cite{Lutchyn2010PRL,Oreg2010PRL,Alicea2010Majorana,Leonid2012NP,Anindya2012NP,MTDeng2012NanoLett,Mourik2012Signatures,
Finck2013PRL,Albrecht2016Nature,Gazibegovic2017Nature,Zhang2018Nature} and the magnetic atomic chain \cite{Stevan2014Science,Jeon2017Science,Michael2017NanoLett} on the surface of an $s$-wave superconductor. Another experimental scheme is the quantum dots chain coupled to an $s$-wave superconductor in a two-dimensional electron gas with spin-orbit coupling and external magnetic field \cite{Fulga2013NJP}. And the quantum dots chain can be viewed as an effective Kitaev chain model under certain conditions. The Anderson-type disorder can be obtained experimentally by controlling the gate potential of the quantum dots \cite{Fulga2013NJP}. Consequently, we believe that the experimental realization of the dimerized Kitaev chain with disorder is promising in the above experimental schemes, especially in the quantum dots chain system.

\section*{Acknowledgments}
B.Z. was supported by the National Natural Science Foundation of China (Grant No. 11274102), the Program for New Century Excellent Talents in University of Ministry of Education of China (Grant No. NCET-11-0960), and the Specialized Research Fund for the Doctoral Program of Higher Education of China (Grant No. 20134208110001). R.C. and D.-H.X. were supported by the NSFC (Grant No. 11704106). D.-H.X. also acknowledges the financial support of the Chutian Scholars Program in Hubei Province.

\appendix
\renewcommand{\appendixname}{\MakeUppercase{appendix}}

\section{\MakeUppercase{Finite-size effect}}
\label{FSE}

\begin{figure}[hptb]
	\includegraphics[width=8cm]{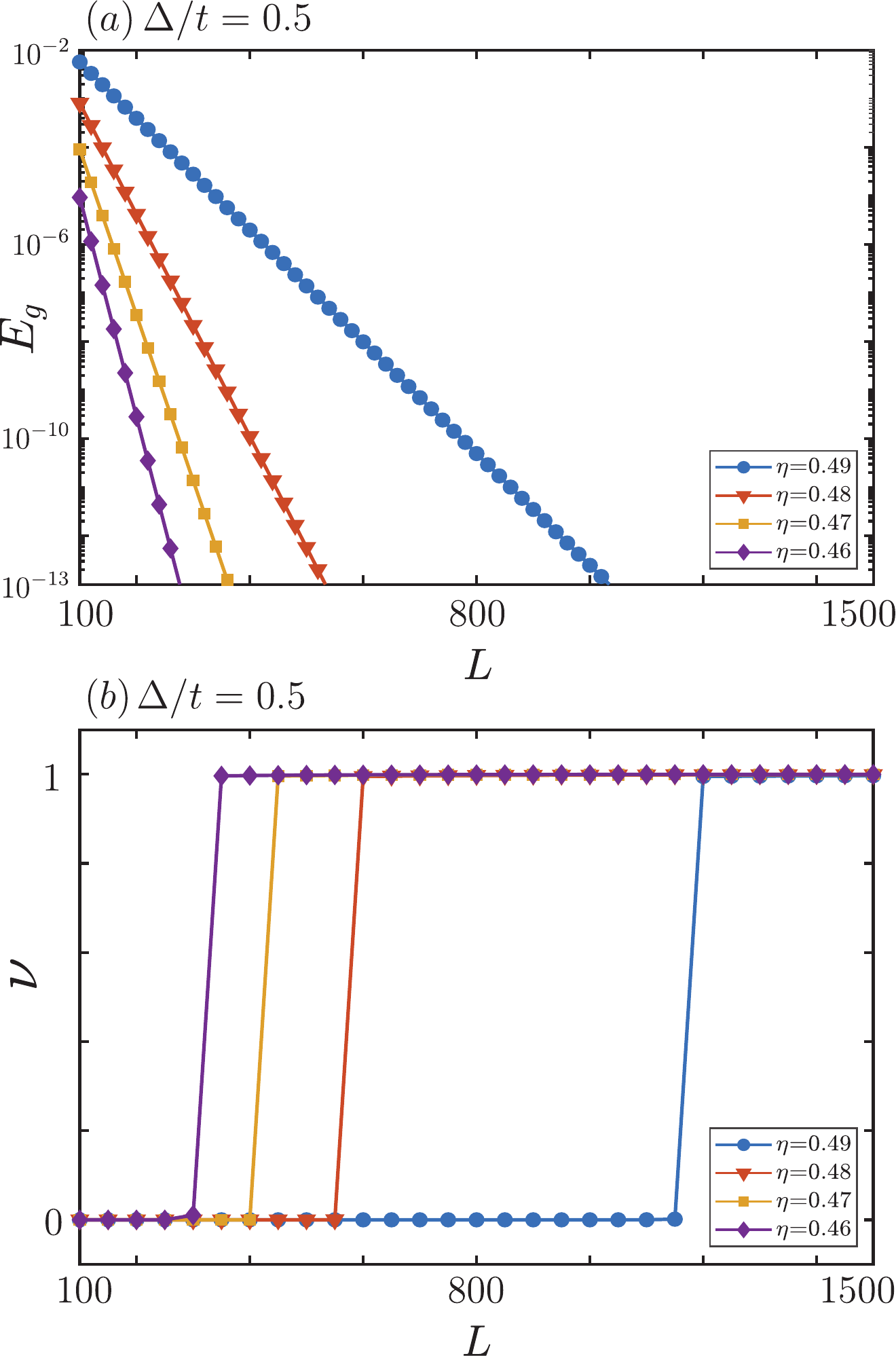} \caption{(Color online)  The finite-size energy gap $E_{g}$ (semilogarithmic plot) and the RSWN $\nu$ of the dimerized Kitaev chain model in the clean limit as a function of the length of the chain for different dimerization parameter $\eta$. We take the parameters $\Delta/t=0.5$ and $\mu=0$. The different dimerization parameter $\eta$ are shown.}%
\label{fig6}
\end{figure}

In the Fig.~\ref{fig4} and~\ref{fig5} of Sec.~\ref{Numerical}, we take the chain length $L=1000$ to avoid the finite size effect \cite{Zhou2011PRB,Chen2019PRB} for numerically calculating the RSWN $\nu$. The finite-size effect refers to that the MZMs on the two ends of the chain can couple together to open an energy gap owing to the finite chain length. Here, we numerically investigate the finite-size effect in the dimerized Kitaev chain model in the clean limit. As shown in Fig.~\ref{fig6}, we plot the finite-size energy gap ($E_{g}$) and the RSWN ($\nu$) as functions of the length of the chain for different dimerization parameter $\eta$. Here, we take the parameters $\Delta/t=0.5$ and $\mu=0$. The parameter value of phase transition is $\eta=0.5$. When the system parameters are selected as $\Delta/t=0.5$ and $|\eta|<0.5$ in Fig.~\ref{fig2}, the corresponding phase is topologically nontrivial phase, and the RSWN $\nu$ should be equal to 1.

Figure~\ref{fig6}(a) display semilogarithmic plot of the finite-size energy gap $E_{g}$ as a function of $L$ for different $\eta$. It is found that $E_{g}$ decays exponentially with increasing chain length $L$, and the decay rates vary with different $\eta$. For the value of $\eta$, the closer the value of phase transition is, the smaller the gap decay rate is. For a given chain length $L$, a larger $\eta$ will give lager $E_{g}$,  and thus the wave function of the end states become more delocalized for $\eta$ approaching $\eta=0.5$ more and more. We also show the RSWN $\nu$ as a function of $L$ for different $\eta$ in Fig.~\ref{fig6}(b). It is found that the very large chain length is required for the numerical calculation of $\nu=1$, when the value of $\eta$ is extremely close to the phase transition value $\eta=0.5$. So in order to present more exactly the topological phase diagram, we choose the chain length $L=1000$.

\section{\MakeUppercase{Derivation of the Self-consistent Born Approximation}}
\label{Derivation}
We start from the one particle Green's function with impurity, and the Green's function is given as \cite{MahanBook,Dniach1998green,Rosenberg2012Phd}
\begin{equation}
G=\left( \omega +i0^{+}-H_{0}-\Sigma \right) ^{-1},
\label{G}
\end{equation}
where $\omega$ is the frequency, $H_{0}$ is the one particle Hamiltonian and $\Sigma$ is the self-energy. To obtain an explicit expression for $\Sigma$, we write it as an infinite summation form of irreducible diagrams \cite{Dniach1998green}. Then, we adopt some approximations. The first approximation is to suppose a low concentration of impurities, and therefore keep only the self-energy diagrams with a single impurity. This means that the impurity scattered by electrons is not simultaneously scattering off another impurity. That is to say, the concentration $n_{i}$ has only the first order term in this sum. The concentration is expressed as $n_{i}\equiv N_{i}/V$, where $N_{i}$ is the number of impurity particles and $V$ is the volume of the system.

The second approximation is that the scattering potential is weak. So that at most two electrons scatter off a single impurity simultaneously, and only the first and second diagrams need to be counted in the sum. The first-order diagram is a line with a single electron and a single impurity, and the second-order diagram is a triangle with two electrons and a single impurity.

The contribution of the first-order diagram to self-energy can be expressed as
\begin{align}
\Sigma ^{1}&= N_{i}U\left( \mathbf{k}=0 \right) \nonumber\\
&= N_{i}\left( \frac{1}{V}\int U\left( \mathbf{x}\right) d^{d}x\right)= n_{i}\int U\left( \mathbf{x}\right) d^{d}x,
\end{align}
where $U$ is the interaction of electron and impurity, and $d$ is the dimension of system. $\Sigma ^{1}$ simply shifts the energies as a constant, and the average is zero. The second-order diagram gives
\begin{equation}
\Sigma ^{2}=N_{i}\sum_{\mathbf{k}^{\prime }}U\left( \mathbf{k}-\mathbf{k}^{\prime
}\right) G\left( \mathbf{k}^{\prime }\right) U\left( \mathbf{k}-\mathbf{k}%
^{\prime }\right).
\label{S2}
\end{equation}
Then, the self-energy in the self-consistent Born approximation is
\begin{equation}
\Sigma\simeq\Sigma ^{1}+\Sigma ^{2}=\Sigma ^{2}.
\label{S3}
\end{equation}

In the dimerized Kitaev chain model investigated in the main text, one need to consider the particle-hole and the sublattice degrees of freedom. For the Anderson-type impurity, the interaction is a delta function $U\left(x\right) =u\left(\tau _{z}\otimes\sigma_{0}\right)\delta\left(x\right)$, where $u$ is the impurity strength, and $\tau_{z}$ and $\sigma_{0}$ are the Pauli matrices respectively acting on the particle-hole and the sublattice degrees of freedom. By Fourier transform, we obtain
\begin{align}
U\left(k\right) &=\frac{1}{V}\int_{-\infty }^{+\infty }U\left(x\right) e^{-i2\pi kx}dx \nonumber \\
&=\frac{1}{V}\int_{-\infty}^{+\infty }u\left(\tau _{z}\otimes\sigma_{0}\right)\delta\left(x\right) e^{-i2\pi kx}dx  \nonumber\\
&=u\left(\tau _{z}\otimes\sigma_{0}\right).
\label{U}
\end{align}
Then, we substitute the Eqs.~(\ref{G}), (\ref{S2}) and (\ref{U}) into the Eq.~(\ref{S3}), and the average strength of the impurities is $\left<u^{2}\right>=\frac{1}{N_{i}W}\int_{-W/2}^{W/2}u^{2}du=\frac{W^{2}}{12N_{i}}$. We transform the sum into the integral form, and the integral can be calculated in the FBZ for a periodic system. Therefore, the self-energy in the self-consistent Born approximation method can be written as
\begin{align}
\Sigma=&\frac{W^{2}}{12}\frac{1}{2\pi }\int_{\text{FBZ}}dk\left( \tau _{z}\otimes \sigma _{0}\right) \nonumber\\
&\times\left[ \omega +i0^{+}-H_{0}\left( k\right) -\Sigma\right] ^{-1}\left( \tau _{z}\otimes \sigma _{0}\right).
\end{align}

\end{document}